\documentclass[aps,prd, twocolumn]{revtex4-1}
\usepackage{amsmath, amssymb, graphicx, color, tikz, hyperref}

\def\realnsig{3.2}

\def\snratio{1.2\%}
\def\baoratio{4.8\%}
\def\mockratio{0.28\%}

\def\aap{Astronomy \& Astrophysics}
\def\aj{Astronomical Journal}
\def\apjl{Astrophysical Journal Letters}
\def\mnras{MNRAS}
\def\jcap{JCAP}
\def\nar{New Astronomy Reviews}
\def\figwidth{0.47\textwidth}

\begin{document}
\title{Key drivers of the preference for dynamic dark energy}

%Please check and correct your affiliation below

\author{Zhiqi Huang}
\email{huangzhq25@mail.sysu.edu.cn}
\affiliation{School of Physics and Astronomy, Sun Yat-sen University, 2 Daxue Road, Zhuhai, 519082, China}

\author{Jianqi Liu}
\affiliation{School of Physics and Astronomy, Sun Yat-sen University, 2 Daxue Road, Zhuhai, 519082, China}

\author{Jianfeng Mo}
\affiliation{School of Physics and Astronomy, Sun Yat-sen University, 2 Daxue Road, Zhuhai, 519082, China}

\author{Yan Su}
\affiliation{School of Physics and Astronomy, Sun Yat-sen University, 2 Daxue Road, Zhuhai, 519082, China}

\author{Junchao Wang}
\affiliation{School of Physics and Astronomy, Sun Yat-sen University, 2 Daxue Road, Zhuhai, 519082, China}

\author{Yanhong Yao}
\affiliation{School of Physics and Astronomy, Sun Yat-sen University, 2 Daxue Road, Zhuhai, 519082, China}

\author{Guangyao Yu}
\affiliation{School of Physics and Astronomy, Sun Yat-sen University, 2 Daxue Road, Zhuhai, 519082, China}

\author{Zhengxin Zhu}
\affiliation{School of Physics and Astronomy, Sun Yat-sen University, 2 Daxue Road, Zhuhai, 519082, China}

\author{Zhuoyang Li}
\affiliation{Department of Astronomy, Tsinghua University, Beĳing 100084, China
} %please add department and address

\author{Zhenjie Liu}
\affiliation{Department of Astronomy, Shanghai Jiaotong University, Shanghai, 200240, China} %please add department and address

\author{Haitao Miao}
\affiliation{National Astronomical Observatories, Chinese Academy of Sciences, Beijing, 100101, China} %please add address

\author{Hui Tong}
\affiliation{School of Physics and Astronomy, Monash University, Vic 3800, Australia}
\affiliation{OzGrav: The ARC Centre of Excellence for Gravitational Wave Discovery, Clayton VIC 3800, Australia}

\date{\today}

\begin{abstract}

  Joint analysis of the baryon acoustic oscillations (BAO) measurement by the Dark Energy Spectroscopic Instrument (DESI) first data release, Type Ia supernovae (SNe) of the Dark Energy Survey Year 5 (DES5YR) release and cosmic microwave background (CMB) data favors a quintom-like dynamic dark energy model over the standard Lambda cold dark matter ($\Lambda$CDM) model at $3.9\sigma$ level (Adame et al. 2024). We confirm the previous finding in the literature that the preference for dynamic dark energy does not rely on the detailed modeling of CMB physics and remains at a similar significance level ($3.2\sigma$) when the full CMB likelihood is replaced by a CMB acoustic-oscillation angle ($\theta_\star$) prior and a baryon abundance ($\Omega_bh^2$) prior. The computationally efficient $\theta_\star$ and $\Omega_bh^2$ priors allow us to take a frequentist approach by comparing DES5YR SNe and DESI BAO with a large number ($\gtrsim 10^4$) of Planck-constrained $\Lambda$CDM simulations. We find that $\geq 3.2\sigma$ preference for dynamic dark energy is very rare (occurrence rate = $0.28\%$) in simulations. When we combine DESI BAO with SN simulations or combine DES5YR SNe with BAO simulations, the occurrence rate of $\geq 3.2\sigma$ preference for dynamic dark energy increases to $1.2\%$ and $4.8\%$, respectively. These results indicate an internal inconsistency, i.e., a significant tension between DESI BAO + DES5YR SNe and Planck-constrained $\Lambda$CDM models in both Bayesian and frequentist points of view. Although both DESI BAO and DES5YR SNe contribute to the preference for dynamic dark energy, the contribution from DES5YR SNe is more significant. In the frequentist point of view, even DES5YR SNe alone is in tension with Planck-constrained $\Lambda$CDM models, though in Bayesian point of view this tension is prior dependent and inconclusive.
\end{abstract}

\maketitle

\section{Introduction \label{sec:intro}}

The past quarter century has witnessed great success of the standard Lambda cold dark matter ($\Lambda$CDM) model, where $\Lambda$ stands for the cosmological constant that acts as dark energy (DE) driving the late-time cosmic acceleration.  The six-parameter $\Lambda$CDM model has been confronted with, and passed the tests of a broad range of cosmological measurements that contain billions of bits of information~\cite{Riess_observational_1998, Eisenstein_detection_2005, Raghunathan_detection_2019, Aghanim_Planck_2020, Alam_completed_2021, Hou_completed_2021, Abbott_DES_2022, Madhavacheril_Atacama_2024}. The cosmic microwave background (CMB) temperature and polarization anisotropies measured by Planck space mission~\cite{Aghanim_Planck_2020} and ground-based telescopes~\cite{Madhavacheril_Atacama_2024, Raghunathan_detection_2019} have determined the $\Lambda$CDM parameters to percentage-level accuracy, ushering in the era of precision cosmology.

As the accuracy of observations continues to improve, however, some data that do not conform to $\Lambda$CDM have emerged~\cite{Riess_comprehensive_2022, Asgari_KIDS_2021, Perivolaropoulos_challenges_2022, Wang_quantifying_2024}. While most studies on cosmological discordance focus on checking self consistency of $\Lambda$CDM without involving an alternative model, some observational data provide hints of dynamic dark energy model that is beyond $\Lambda$CDM~\cite{Zhao_dynamical_2017, Li_cosmological_2018, Sola_dynamical_2018, Wang_Pantheon_2022, Adame_DESI_2024}. The most recent revival of dynamic dark energy comes from the measurement of baryon acoustic oscillations (BAO) by the Dark Energy Spectroscopic Instrument (DESI) first data release~\cite{Adame_DESI_2024}. The DESI BAO data, when combined with CMB and  type Ia supernovae data from Dark Energy Survey Year 5 (DES5YR) release~\cite{Abbott_DES_2024}, favors a dynamic dark energy model with the equation of state (EOS) of dark energy parameterized as $w = w_0+w_a(1-a)$, where $a$ is the scale factor~\cite{Chevallier_accelerating_2001, Linder_exploring_2003}. The joint analysis of DESI BAO + DES5YR SNe + CMB gives $w_0 = -0.727\pm 0.067$ and $w_a = -1.05^{+0.31}_{-0.27}$, which favors a quintom-like ($w$ crossing $-1$) dynamic dark energy scenario~\cite{Feng_dark_2005} and rejects the $\Lambda$CDM model ($w_0=-1$ and $w_a=0$) at $3.9\sigma$ confidence level~\cite{Adame_DESI_2024}. The significance level, however, drops to $3.5\sigma$ and $2.5\sigma$ when DES5YR SNe is replaced by the Union3~\cite{Rubin_Union_2023} and Pantheon+ supernova data~\cite{Brout_Pantheon_2022}, respectively.

The origin of the preference for dynamic dark energy has been investigated in several works~\cite{Adame_DESI_2024, Colgain_DESI_2024, Wang_self_2024, Wang_role_2024}, yet it remains unclear why the joint statistical significance of the preference for dynamic dark energy is sensitive to the choice of supernova data. It has also been shown that the significance level has subtle dependence on the prior range of dark energy parameters and whether Atacama Cosmology Telescope (ACT) data release 6 CMB lensing data is included~\cite{Wang_self_2024}. To address these confusions, the present work aims to identify which data set is the key driver of the preference for dynamic dark energy.

Throughout the paper we work with natural units $c=\hbar=k_B=1$ and a spatially flat Friedmann-Robertson-Walker (FRW) background metric. The Hubble constant $H_0$ is written as $100  h\;\mathrm{km\, s^{-1}\mathrm{Mpc}^{-1}}$, where the dimensionless parameter $h$ is the reduced Hubble constant. The abundance parameters $\Omega_c$, $\Omega_b$, $\Omega_{\rm DE}$, $\Omega_{\gamma}$, $\Omega_{\nu i}$ ($i=1,2, 3$) are defined as the present fractional background density of cold dark matter, ordinary matter, dark energy, radiation, and the $i$-th neutrino, respectively. The matter abundance is defined as $\Omega_m = \Omega_b + \Omega_c$.

\section{Method}

For spatially flat $w_0w_a$CDM cosmology, the Hubble expansion rate $H(z) > 0$ is determined by the first Friedmann equation
\begin{eqnarray}
  \left[\frac{H(z)}{H_0}\right]^2 && = \Omega_m(1+z)^3 + \Omega_{\rm DE}e^{3\left[(1+w_0+w_a)\ln (1+z) - w_a\frac{z}{1+z}\right]} \nonumber \\
  && + \left[\Omega_\gamma + \sum_{i=1}^3 \Omega_{\nu,i}\frac{\mathrm{I}_{\rho}\left(\frac{m_{\nu,i} }{(1+z)T_{\rm CNB}}\right)}{\mathrm{I}_{\rho}\left(\frac{m_{\nu,i}}{T_{\rm CNB}}\right)}\right](1+z)^4, \label{eq:H} 
\end{eqnarray}
where $z$ is the cosmological redshift; $m_{\nu, i}$ is the neutrino mass of the $i$-th specie; $T_{\rm CNB}=T_{\rm CMB}\left(\frac{4}{11}\right)^{1/3}\approx 1.95\,\mathrm{K}$ is the effective temperature for neutrino momentum distribution. The neutrino density integral is
\begin{equation}
  \mathrm{I}_\rho(\lambda) \equiv  \frac{1}{2\pi^2}\int_0^\infty \frac{x^2\sqrt{x^2+\lambda^2}}{e^x+1}\mathrm{d}x.
\end{equation}
As in the standard scenarios $\lesssim 0.1\mathrm{eV}$ neutrino masses have almost no impact on the determination of dark energy parameters~\cite{Aghanim_Planck_2020, Liu_cosmological_2024}, we fix three neutrino species with masses $0.05\mathrm{eV}$, $0.009\mathrm{eV}$, $0.001 \mathrm{eV}$ for simplicity. The abundance parameter $\Omega_{\nu, i}$ is related to the mass parameter $m_i$ via
\begin{equation}
  \Omega_{\nu, i}h^2 = 1.981\times 10^{-5}\mathrm{I}_{\rho}\left(\frac{m_{\nu,i}}{T_{\rm CNB}}\right),
\end{equation}
where a standard CMB temperature $T_{\rm CMB}=2.726\,\mathrm{K}$ effective neutrino number $N_{\rm eff}=3.044$ have been used.
The comoving angular diameter distance is given by
\begin{equation}
  D_M(z) = \int_0^z\frac{dz'}{H(z')}. \label{eq:DM}
\end{equation}
The sound horizon at the baryon drag epoch, $r_d$, is approximated by~\cite{Brieden_tale_2023}
\begin{equation}
  \frac{r_d}{147.05\,\mathrm{Mpc}} = \left(\frac{\Omega_mh^2}{0.1432}\right)^{-0.23}\left(\frac{N_{\rm eff}}{3.04}\right)^{-0.1}\left(\frac{\Omega_bh^2}{0.02236}\right)^{-0.13}. \label{eq:rd}
\end{equation}
The BAO data measures $D_M/r_d$, $1/(Hr_d)$ and  $D_V\equiv (zD_M^2/H)^{1/3}$. We apply multi-variable Gaussian likelihood for the data and covariance matrix listed in Table 1 of DESI 2024 paper~\cite{Adame_DESI_2024}, but with a better decimal precision obtained via private communication.

The distance modulus of SNe  is
\begin{equation}
  \mu(z) = 5\log_{10}\frac{(1+z_{\rm hel})D_M(z_{\rm CMB})}{\mathrm{Mpc}} + 25, 
\end{equation}
where $z_{\rm hel}$ is the heliocentric redshift and $z_{\rm CMB}$ is the CMB dipole-causing peculiar redshift removed. A supernova data set with observed distance moduli $\mu_i^{\rm obs}$ ($i=1,2,\ldots, N$) and their covariance matrix $C$ has a chi-square function (defined as $-2$ times the logarithm of likelihood)
\begin{equation}
  \chi^2 = \sum_{i, j}\left(\mu_i^{\rm obs} - \mu_i^{\rm th} - \Delta\right)\left(C^{-1}\right)_{ij} \left(\mu_j^{\rm obs} - \mu_j^{\rm th} - \Delta\right), \label{eq:SNchisq}
\end{equation}
where $\mu^{\rm th}$ is the theoretical distance modulus and
\begin{equation}
  \Delta = \frac{\sum_{i,j} \left(\mu_i^{\rm obs} - \mu_i^{\rm th}\right)\left(C^{-1}\right)_{ij}}{\sum_{i, j} \left(C^{-1}\right)_{ij}} \label{eq:Delta}
\end{equation}
can be considered as the average difference between observation and theory. Because we are marginalizing over the absolute magnitude of SNe, the average difference $\Delta$ enters Eq.~\eqref{eq:SNchisq} in such a way that the likelihood is invariant under a global translation of $\mu^{\rm th}$. Eq.~\eqref{eq:Delta} can be used to define a local $\Delta_{\rm bin}$ for a subset of SN data in a redshift bin.  The difference $\Delta_{\rm bin} -\Delta$, which we dub ``binned distance modulus'',  measures how the data in the redshift bin deviates from theoretical prediction.

The list of $z_{\rm hel}$, $z_{\rm CMB}$, $\mu^{\rm obs}$ and the covariance matrix $C$ for DES5YR SNe data are all publicly available at \url{https://github.com/des-science/DES-SN5YR/}.

Due to strong degeneracy between parameters, a single probe of SNe, BAO or CMB cannot effectively constrain the $w_0w_a$CDM model. Joint analysis with CMB + BAO + SNe breaks the degeneracy and gives a prior-insensitive constraint~\cite{Adame_DESI_2024, Wang_self_2024}. In this way, the constraint on dark energy parameters involves CMB physics and relies on many details of the cosmological model. The robustness of dark-energy constraint can be vastly enhanced if we treat CMB as a BAO data point at recombination (redshift $z\sim 1090$) and only use the constraint on the angular extension of the sound horizon on the last scattering surface, i.e., the $\theta_\star$ parameter. The CMB constraint on $\theta_\star$ is only sensitive to the spatial curvature and the early-universe physics, and is almost model-independent if we work with the standard scenario of Big Bang Nucleosynthesis (BBN) in a spatially flat universe. Based on the Planck constraint~\cite{Aghanim_Planck_2020}, we adopt
\begin{equation}
  \theta_\star = 0.0104092 \pm 0.0000032. \label{eq:theta}
\end{equation}

To use the CMB and BAO standard ruler information, it is necessary to determine the sound horizon at the recombination epoch, which depends on the baryon abundance $\Omega_bh^2$. We quote a BBN constraint~\cite{Schoneberg_2024_2024}
\begin{equation}
  \Omega_b h^2 = 0.02196\pm 0.00063. \label{eq:ombh2}
\end{equation}
 We do not use a tighter $\Omega_bh^2$ prior from CMB, because it relies on slightly more sophisticated modeling of CMB and slightly depends on the dark energy model under investigation. Hereafter we denote the CMB $\theta_\star$ and BBN $\Omega_bh^2$ constraints simply as ``$\theta_\star$'' when brevity is needed.

With the DESI BAO and DES5YR SNe likelihoods and constraints on $\theta_\star$ and $\Omega_bh^2$, we run Markov Chain Monte Carlo (MCMC) simulations to measure the cosmological parameters that are listed in Table~\ref{tab:params}.  

\begin{table}
  \caption{cosmological parameters \label{tab:params}}
  \begin{tabular}{lll}
    \hline
    \hline
    parameter & definition & prior range \\
    \hline
    $\Omega_m$ & matter abundance & $[0.05, 0.8]$ \\
    $w_0$ & dark energy EOS at $z=0$ & $[-3, 1]$ \\
    $w_a$ & dark energy EOS slope & $[-3, 3]$ \\
    $h$ & reduced Hubble constant & $[0.5, 0.9]$ \\
    $\Omega_bh^2$ & baryon density & $[0.015, 0.03]$ \\
    \hline
  \end{tabular}
\end{table}

With the combination $\theta_\star$+BAO+SN, the marginalized posterior of $(w_0, w_a)$ is well approximated by a two dimensional Gaussian distribution. For each simulation, we run a long Markov chain ($\gtrsim 10^4$ samples after burn in) and marginalize over other parameters to obtain well converged mean and covariance matrix for $(w_0, w_a)$, denoted as $(\bar{w}_0, \bar{w}_a)$ and $\mathrm{Cov}(w_0, w_a)$, respectively. We define the significance level of the preference for dynamic dark energy by the smallness of the tail probability of $\Lambda$CDM, i.e. $(w_0=-1, w_a=0)$, in the Gaussian approximation. The tail probability can be converted to the number of sigmas ($n_\sigma$) via  
\begin{equation}
\mathrm{erf}\,\left(\frac{n_\sigma}{\sqrt{2}}\right) = \exp{\left[-(1+\bar{w}_0, \bar{w}_a) \mathrm{Cov}^{-1}(w_0, w_a) \begin{pmatrix} 1+\bar{w}_0 \\ \bar{w}_a\end{pmatrix}\right]}. \label{eq:nsig}
\end{equation}

To test how likely the preference for dynamic dark energy is due to statistical fluctuations, we take a frequentist approach and compare the DESI BAO and DES5YR SNe data with Planck-constrained $\Lambda$CDM simulations. For each simulation we randomly draw $\Lambda$CDM parameters from the Planck posterior, which has been obtained by running MCMC with Planck TTTEEE+lowl+lowE+lensing likelihood and flat priors on $(\Omega_bh^2, \Omega_ch^2, \theta_\star, \ln A_s, n_s)$~\cite{Aghanim_Planck_2020}.

\section{Results}

\begin{figure}
  \vspace{0.05in}
  \includegraphics[width=\figwidth]{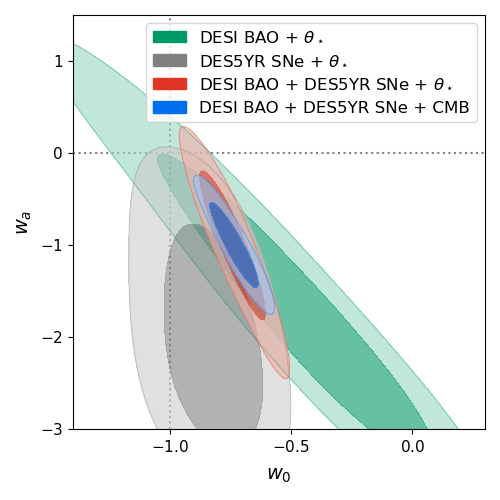}
  \caption{Marginalized 68.3\% and 95.4\% confidence constraints on $w_0$ and $w_a$. \label{fig:w0wa}}
\end{figure}

We first show the results of a standard Bayesian analysis for the parameters and their priors given in Table~\ref{tab:params}. As Figure~\ref{fig:w0wa} shows,  BAO+$\theta_\star$ (green contours) is well consistent with $\Lambda$CDM, and SNe+$\theta_\star$ (gray contours) is marginally consistent ($\sim 2\sigma$) with $\Lambda$CDM. These constraints, however, as pointed out in~Ref.~\cite{Adame_DESI_2024}, are prior-dependent. In particular, if we change the prior range of $w_a$ from $[-3,3]$ to $[-5, 5]$, SNe+$\theta_\star$ disfavors  $\Lambda$CDM at $2.9\sigma$ level. Further broadening the $w_a$ prior range leads to even stronger tension between $\Lambda$CDM and DES5YR SNe + $\theta_\star$.

The prior-dependence becomes negligible when we combine DESI BAO, DES5YR SNe and $\theta_\star$ all together (red contours). In this case we obtain $w_0=-0.756^{+0.089}_{-0.087}$ and $w_a = -0.904^{+0.489}_{-0.494}$, excluding $\Lambda$CDM model at $\realnsig\sigma$ significance level. For a comparison between $\theta_\star$ and full CMB likelihood, we replace $\theta_\star$ with the full Planck TTTEEE+lowl+lowE+lensing likelihood and obtain the constraint shown as blue contours, which have been similarly shown in Figure 6 of Ref.~\cite{Adame_DESI_2024}. Despite losing a bit constraining power, the $\theta_\star$ approximation captures the major feature of CMB constraint and is much more computationally efficient. Therefore, for running many MCMC simulations with over $10^4$ realizations of mock data, we use $\theta_\star$ instead of the full CMB likelihood.

While nothing precludes using the particular combination of DESI BAO and DES5YR SNe and confront it with the particular $w_0w_a$CDM model, given that there is yet no independent confirmation from other cosmological probes, it is reasonable to speculate unknown systematic errors in either DESI BAO or DES5YR SNe~\cite{Ding_theoretical_2018, Chen_baryon_2024, Wang_revisiting_2023, Yao_progenitor_2024, Efstathiou_evolving_2024}. When Planck-constrained $\Lambda$CDM is assumed, such speculation is in fact testable. If the preference for dynamic dark energy is driven by, for instance, some unknown systematics in DESI BAO, we can replace DES5YR SNe with Planck $\Lambda$CDM mocks at the same redshifts and with the same error matrix. There should be some chance that DESI BAO + SN mock + $\theta_\star$ favors $w_0w_a$CDM over $\Lambda$CDM at a significance level that is equal to or greater than the real-data case. Similarly if some unknown systematics in DES5YR SNe is at play, there should be some chance that DES5YR SNe + BAO mock + $\theta_\star$ favors $w_0w_a$CDM over $\Lambda$CDM at an equal or greater significance level. 

\begin{table}
  \caption{data sets and occurrence rate of $n_\sigma \geq \realnsig$\label{tab:hyp}}
  \begin{tabular}{cc}
    \hline
    \hline
    MCMC data sets & P($n_\sigma \geq \realnsig$) \\
    \hline
     BAO mock + SN mock + $\theta_\star$ & $\mockratio$ \\
    \hline
    DESI BAO + SN mock + $\theta_\star$   & $\snratio$ \\
    \hline
    BAO mock + DES5YR SNe +  $\theta_\star$  & $\baoratio$ \\
    \hline
  \end{tabular}
\end{table}

\begin{figure}
  \includegraphics[width=\figwidth]{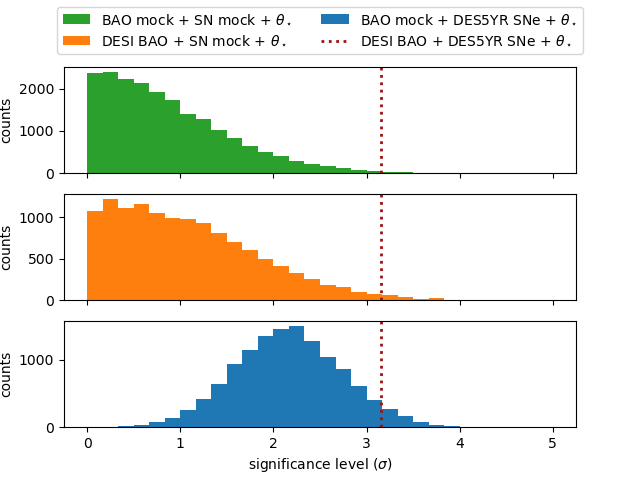}    
  \caption{Histogram of significance level of favoring $w_0w_a$CDM over $\Lambda$CDM, for SN mock + BAO mock + $\theta_\star$ (upper panel), DESI BAO + SN mock + $\theta_\star$ (middle panel),  and DES5YR SNe + BAO mock + $\theta_\star$ (lower panel),  respectively. The dotted vertical line ($\realnsig\sigma$) is the significance level from the real data (DESI BAO + DES5YR SNe + $\theta_\star$).  \label{fig:nsig}}
\end{figure}

We then perform a frequentist analysis by replacing DESI BAO and/or DES5YR SN with Planck-constrained $\Lambda$CDM simulations and observe the occurrence rate of $n_\sigma \ge \realnsig$, i.e., how likely the $\geq \realnsig\sigma$ preference for dynamic dark energy can arise from statistical fluctuations. Table~\ref{tab:hyp} lists the data sets used in MCMC runs and $P(n_\sigma \geq \realnsig)$ that denotes the occurrence rate of getting $\realnsig$ preference for dynamic dark energy. For each row we generate more than $10^4$ simulations with different random seeds, and run MCMC for each simulation. Figure~\ref{fig:nsig} shows the histogram of $n_\sigma$ from the MCMC runs.

In the first case where SN mocks and BAO mocks are used, the histogram of $n_\sigma$ (upper panel of Figure~\ref{fig:nsig}) approximately follows a Gaussian distribution ($\sim e^{-n_\sigma^2/2}$ for $n_\sigma\geq 0$),  confirming the agreement between Bayesian and frequentist approaches. The very low occurrence rate ($\mockratio$) of $n_\sigma\geq \realnsig$ suggests that in a Planck-constrained $\Lambda$CDM model the preference for dynamic dark energy can hardly be interpreted as a statistical fluctuation, or equivalently, that there is a significant tension between DESI BAO + DES5YR SNe and Planck-constrained $\Lambda$CDM models.

In the second case where DESI BAO instead of BAO mocks is used,  the histogram of $n_\sigma$ (middle panel of Figure~\ref{fig:nsig}) slightly shifts towards larger values of $n_\sigma$. The occurrence rate of $n_\sigma\ge \realnsig $ increases to $\snratio$, indicating a noticeable contribution from DESI BAO to the preference for dynamic dark energy. However, since the distribution peak is still around $n_\sigma = 0$, the DESI BAO data alone is still consistent with Planck-constrained $\Lambda$CDM. This is in agreement with the Bayesian result of fitting $w_0w_a$CDM to DESI BAO + $\theta_\star$ (green contours in Figure~\ref{fig:w0wa}).

In the final case where DES5YR SNe instead of SN mocks are used, the histogram of $n_\sigma$ (lower panel of Figure~\ref{fig:nsig}) significantly shifts towards larger values of $n_\sigma$. The occurrence rate of $n_\sigma\ge \realnsig$ increases to $\baoratio$, indicating a more significant contribution from DES5YR SNe to the preference for dynamic dark energy. Moreover, the exclusion of $n_\sigma=0$ in this case indicates internal inconsistency, i.e., a tension between DES5YR SNe and Planck-constrained $\Lambda$CDM models. In the previous Bayesian analysis of fitting $w_0w_a$CDM to DES5YR SNe + $\theta_\star$, the tension between $\Lambda$CDM and DES5YR SNe is prior-dependent,  $\sim 2\sigma$ for $w_a\in[-3,3]$ and $\sim 3\sigma$ for $w_a\in[-5, 5]$. Here the frequentist analysis tends to agree with the Bayesian analysis with a broader $w_a$ prior range.

\begin{figure}
  \includegraphics[width=\figwidth]{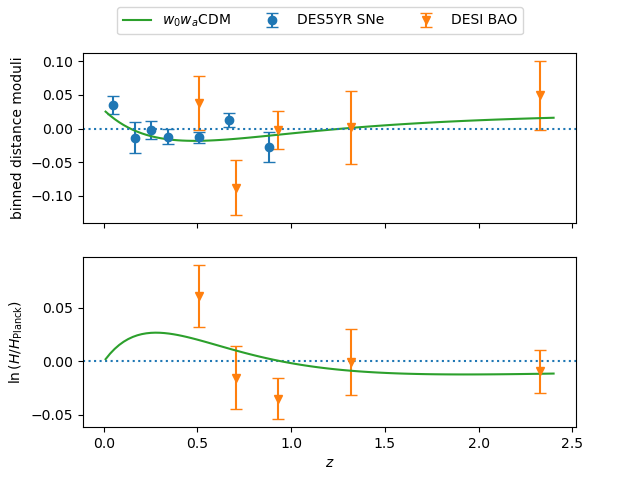}
  \caption{Binned distance moduli and $\ln H$. The Planck best-fit $\Lambda$CDM is subtracted. The green solid line is the prediction of the DES5YR SNe + DESI BAO + $\theta_\star$ best-fit $w_0w_a$CDM model, where $w_0=-0.76$, $w_a=-0.88$, and $\Omega_m=0.314$. \label{fig:data}}
\end{figure}

Since both DESI BAO and DES5YR SNe contribute to the preference for dynamic dark energy, it is instructive to compare the two data set in a same observable space and see if there is any signal in common. If we take $r_dh$ as a free parameter and marginalize over it, the BAO measurements of $D_M/r_d$ (transverse BAO information) can be translated to constraints on distance moduli in the same form of SN data. The upper panel of Figure~\ref{fig:data} shows the binned distance moduli of SNe and BAO, with the Planck best-fit $\Lambda$CDM prediction subtracted. We do not find similar patterns in the distance-modulus chart for DES5YR SNe and DESI BAO. While the DES5YR SNe distance-modulus data clearly follow the prediction of DES5YR SNe + DESI BAO + $\theta_\star$ best-fit $w_0w_a$CDM model, the DESI BAO $D_M/r_d$ data do not. Only when including the radial BAO information (the $H$ data in the lower panel of Figure~\ref{fig:data}), the full DESI data set shows a weak ($< 2\sigma$) preference for dynamic dark energy, as indicated by both Bayesian and frequentist analyses.

%What pushes an overall fit to $w_0w_a$CDM in the DESI BAO data is the anomalously  large $H$ at effective redshift $z_{\rm eff}=0.51$~\cite{Adame_DESI_2024, Colgain_DESI_2024}, as shown in the lower panel of Figure~\ref{fig:data}, as well as the anomalously low $D_M$ at effective redshift $z_{\rm eff}=0.71$~\cite{Wang_role_2024} shown in the upper panel of Figure~\ref{fig:data}.

\section{Discussion and Conclusions}

In this work, we demonstrate that the recently claimed preference for dynamic dark energy in DESI+DES5YR+CMB does not depend on the physical details of CMB. Replacing the full CMB information with a simple and model-independent $\theta_\star$ constraint leads to $\realnsig\sigma$ exclusion of $\Lambda$CDM in $w_0w_a$CDM cosmology. This is in agreement with similar findings in, e.g., Ref.~\cite{Adame_DESI_2024}. To test how likely the $\realnsig\sigma$ exclusion of  $\Lambda$CDM can be due to statistical fluctuations, we take a frequentist approach by comparing the real data with $\gtrsim 10^4$ Planck-constrained $\Lambda$CDM simulations. We find the preference for dynamic dark energy is unlikely ($\sim 0.28\%$ chance to be) a statistical fluctuation, if Planck-constrained $\Lambda$CDM is the correct cosmology. Further analysis show that both DESI BAO and DES5YR SNe contribute to the preference for dynamic dark energy, and the contribution from DES5YR SNe is more significant. In the frequentist point of view, DES5YR SNe alone is already in tension with Planck-constrained $\Lambda$CDM models, though in Bayesian point of view the tension between DES5YR SNe and $\Lambda$CDM is prior-dependent and hence inclusive.

Throughout the work we have compared the data against Planck-constrained $\Lambda$CDM cosmology. We would like to point out that the prior on cosmology can have a significant impact on the result. For instance, if we made a stronger assumption by fixing the cosmology to Planck best-fit $\Lambda$CDM instead of  randomly drawing $\Lambda$CDM parameters from Planck posterior, we would get stronger result, i.e., smaller occurrence rates of $\geq \realnsig$ preference for dynamic dark energy ($0.9\%$ for $\theta_\star$ + DES5YR SNe + BAO mocks and $2.1\%$ for $\theta_\star$ + DESI BAO + SN mocks). It is conceivable that if we use a weaker prior on $\Lambda$CDM parameters, our result would be weakened.

Although both DESI BAO and DES5YR SNe have a trend of favoring $w_0w_a$CDM over $\Lambda$CDM, the signals are from different observable spaces. The transverse BAO information ($D_M/r_d$ data) does not have a clear trend to follow the pattern of the binned distance moduli of DES5YR SNe. We also check that a different SN data set Pantheon+ does not follow this pattern, neither. It is therefore yet too early to claim that the data consistently point to the dynamic dark energy model. 

\section{Acknowledgements}

This work is supported by the National Natural Science Foundation of China (NSFC) under Grant No. 12073088, National key R\&D Program of China (Grant No. 2020YFC2201600), and National SKA Program of China No. 2020SKA0110402. The computing resources are supported by Weide Liang, Mingcheng Zhu, Ren-Peng Zhou,  Zhenyang Huang, Penghui Dai and Huan Zhou via the MEET-U (``Make Everyone Explore The Universe'') outreach website~(\url{http://zhiqihuang.top/MEETU}), where we make the simulation source code publicly available and allow astronomy amateurs to contribute their computing resources. We are grateful to Kyle Dawson for kindly sharing the data used in the DESI 2024 release. We owe a debt of gratitude to the anonymous referee who has made very insightful comments that greatly improved the quality of this work.

%\bibliographystyle{apsrev4-1}
%\bibliography{ref}

%merlin.mbs apsrev4-1.bst 2010-07-25 4.21a (PWD, AO, DPC) hacked
%Control: key (0)
%Control: author (72) initials jnrlst
%Control: editor formatted (1) identically to author
%Control: production of article title (-1) disabled
%Control: page (0) single
%Control: year (1) truncated
%Control: production of eprint (0) enabled
%

\end{document}